# Electronic Properties of Various Graphene Quantum Dot Structures: an Ab Initio Study

M. Ghandchi[1], G. Darvish[*,2], M. K. Moravvej-Farshi[3]

1- Department of Electronics, Mechanics, Electrical and Computer Engineering Faculty, Science and Research Branch, Islamic Azad University, Tehran, Iran, majid.ghandchi@srbiau.ac.ir
2- Department of Electronics, Mechanics, Electrical and Computer Engineering Faculty, Science and Research Branch, Islamic Azad University, Tehran, Iran, darvish_gh@srbiau.ac.ir
3- Department of Electronics, Faculty of Electrical and Computer Engineering, Nano-Plasmo Photonic Research Group, Tarbiat Modares University, Tehran, Iran, moravvej@modares.ac.ir

[*]Corresponding author,

**Abstract**
   Density functional theory (DFT) and thermal DFT (thDFT) calculations were used to evaluate the energy band structure, bandgap, and the total energy of various graphene quantum dots (GQDs). The DFT calculations were performed using local density approximation for the exchange-correlation functional and norm-conserving pseudopotentials. We consider the triangular and hexagonal GQDs with zigzag and armchair edges and 1-3 nm dimensions with many hundred atoms. The simulation results show that all of these GQDs are direct bandgap semiconductors with a flat band structure, and they are suitable for electronics and optoelectronics applications. Analysis of GQDs in which the A and B sublattice symmetries were broken showed degenerate zero-energy shells. Using the thDFT calculations carried out at temperatures up to 1400 K, we evaluated the temperature dependence of the GQDs bandgaps and total energies via entropy-term and electron's kinetic energy. The obtained results indicate that the ground-state DFT calculations are valid for determining the electronic properties of GQDs up to room temperature. Moreover, we tune semi-empirical parameters of the tight-binding model by the DFT results in small GQDs to reduce the computational cost of electronic structure calculations for large GQDs, which contained up to thousands of atoms.

## 1. Introduction

As a newly emerging material, graphene is highly important in electronic and optoelectronic devices [1, 2]. Thanks to its unique properties in tuning by applying voltage and changing its chemical potential, graphene is used to design and construct fields and waves devices, even in telecommunication applications [3].

The zero bandgap ($E_g$) of graphene enables the design of electronic materials using various techniques to create a bandgap. In this regard, graphene quantum dots (GQDs) are considered one of the most efficient engineering methods for creating an energy gap in graphene [4]. The quantum dots technique has already been used for III-V semiconductors. This technique causes an energy level discretization in their electronic structure [5]. Thus, in graphene, the necessary changes can be made on the energy bands with the quantum dots' technique. Compared to graphene nanoribbons (GNRs) [6-8], GQDs allows engineering of the geometry, size, and edge type of GQDs, including circular, triangular, and hexagonal quantum dots [9-12].

When selecting GQDs for the design of a device, the first issue with which the engineers are faced is to obtain information about their electronic structures and energy bands. Then, the density of states (DOS) is obtained, paving the way for the calculating of the optical absorption spectrum and carrier transport calculations so that $I$-$V$ characteristics of the GQDs can be obtained.

Various valuable studies have reported the investigation of the electronic properties of graphene and nano-graphene. The approaches of these studies can be divided into several categories. The tight-binding (TB) model is one of the most common tools for calculating the electronic properties of nano-graphene [13]. The TB model approach was used to study GQDs by other researchers, for example, in the study by Zarenial et al. [14]. Density functional theory (DFT) is another method that has been applied for the study of electronic properties [15, 16]. There is no need for any experimental parameters in this method, and the calculations are carried out solely based on the molecular-atomic structure and first-principles, obtaining highly accurate results. The use of first-principles calculations, which do not require tuning semi-empirical coefficients to perform simulations, is a reliable method and precedes experimental studies. Thus, the DFT method is an acceptable and appropriate computational method for nano-graphene and other emerging nanomaterials such as Germanene [17], which have quantum confinement. Investigations of the electronic properties of GQDs based on DFT have already been reported in the literature [18-20]. Among the DFT method's constraints, the need for extensive computer resources for solving the Kohn-Sham equations can be pointed out. Another limitation of DFT includes applying the independent particle approximation (IPA) and the lack of consideration of the electron-electron (e-e) interaction. Thus, in the third approach, electronic properties' studies are carried out based on the multiparticle model (many-body), and the e-e interaction is also considered [21, 22].

This article investigates the electronic properties of GQDs using ab initio study with two methods, namely, standard DFT and thermal DFT (thDFT). All parameters of GQD, such as size, shape, edge type, and A and B sublattice symmetry (degenerate states), are considered. We consider the triangular and hexagonal GQDs with zigzag and armchair edges and 1-3 nm dimensions with many hundred atoms. In degenerate GQDs, the bandgap calculation rules are discussed, and their differences from the nondegenerate GQDs are declared. Another notable point in our electronic properties calculations highlights the temperature dependence in GQDs via entropy-term and electron's kinetic energy (up to 1400 K). We also compare the DFT results with the TB model outputs to tune the semi-empirical parameters of the TB model using the DFT calculations results. Thus, such a tuned TB model can simulate GQDs with many atoms using fewer computation resources while achieving good accuracy that is similar to that of the DFT method.

## 2. Computational methods

The electronic properties of various GQDs structures are investigated using the DFT method. The side edges of all of the GQD systems are passivated with hydrogen atoms to avoid dangling bonds. Additionally, finite system conditions are used to perform DFT calculations, and each of the GQDs is considered an isolated system. For this purpose, we have taken advantage of a rectangular cubic vacuum layer that surrounds our entire atomic structure. In isolation systems, we use periodic boundary conditions (PBC) to perform DFT calculations. We put GQD in a big box, and there is enough space around it. The open-source Quantum ESPRESSO package [23, 24] is used to implement DFT calculations. The use of QUANTUM ESPRESSO software in the field of Nano-electronics has increased growingly, and a significant number of researchers use this open-source software package to perform quantum computations containing Schrödinger equations [17]. We use the LDA functional [25] and the NCPS pseudopotential [26, 27] with this software. By examining the previous calculations performed by other researchers using the tight-binding model [28], we select the desired structures among the GQDs to achieve the best configuration for optoelectronics and photonics applications. Figure 1 illustrates a simple schematic of the different GQD structures. For each case, we examine different sizes of quantum dots. The JSME software packages [29] are used to design and arrange atomic structures. In addition, the Xcrysden [30] and VESTA [31, 32] software packages are used for 3D visualization and to investigate the geometric accuracy of the designed structures. This geometric investigation is performed prior to starting the computationally intensive DFT calculations on GQDs. We use a totally-integrated set of open-source software tools in this project. Therefore, one feature of our work in this research is useing of the Linux operating system and computational codes based on general public license GNU[1].

It should be noted that we used the "*minimum energy configuration*" [33] before the DFT calculation step. This configuration was done to obtain the proper (optimum) values of the computational parameters, including the lattice constant, cut-off energy, and vacuum layer thickness. Furthermore, we investigate the temperature dependence of the bandgap and the total energy of the GQD structures using the thDFT computational method. In thDFT, the Mermin-Kohn-Sham equations are solved integrally [34], and the electron temperature is used with the Fermi-Dirac distribution to obtain the occupation function of the electronic states. In standard DFT calculations, the charge density $\rho(r)$ is derived from (1):

$$\rho(r) = \sum_i^N |\phi_i(r)|^2 \quad (1)$$

where $\phi_i(r)$ is the $i^{th}$ Kohn-Sham orbital. However, for thDFT, the temperature-dependent charge density $\rho^\tau(r)$ is obtained by considering the Fermi-Dirac distribution function as follows:

$$\rho^\tau(r) = \sum_i^N f_i |\phi_i^\tau(r)|^2 \quad (2)$$

where $f_i$ is the Fermi occupation factor that is defined by:

$$f_i = \frac{1}{1+\exp(\frac{\varepsilon_i^\tau - \mu}{\tau})} \quad (3)$$

where $\mu$ is the chemical potential, $\varepsilon_i^\tau$ is the total energy of an electron located in the $i^{th}$ orbital with $\tau$ thermal energy component, that $\tau = k_B T$, and where $k_B$ is the Boltzmann constant and $T$ is the electron temperature in Kelvin. Also, by a functional rewriting of computation of xc energy in DFT calculations, it is possible to make more exact temperature dependence. For example, the thermal LDA (thLDA) functional can be created [35, 36], which can be defined as an independent study. We have used the LDA functional in this study [37]. However, recent studies conducted by other researchers confirm [34] that the effect of using the thLDA functional on the final results of electronic properties computations at high temperatures is negligible, and the LDA-based results in the present work are sufficient for most application areas. Here thermal DFT calculations are performed using the NanoDCAL software package [38].

---

[1] By Stallman, Richard, 1998.

## 3. Results and discussion
### 3.1. Electronic bands structure

In the "*minimum energy configuration*," it is necessary to repeatedly carry out the DFT calculations until the calculator software's optimum parameters for achieving the minimum total energy are obtained. Figure 2 shows the effect of the vacuum layer thickness and cut-off energy parameters on the DFT calculation results. It is observed from Fig. 2 that by adopting cut-off energy of 40 Ry, the obtained total energy will reach its minimum value and will no longer be reduced by increasing the cut-off energy. Furthermore, the vacuum layer thickness's effect on the total energy is illustrated in Fig. 2. As observed from the figure, the change in the vacuum layer thickness parameter does not significantly affect the DFT results.

Figure 3 shows the electronic properties of the $C_{42}acH_{18}$ hexagonal GQD with armchair edges, including the band structure, molecular energy spectrum (MES), bandgap ($E_g$), density of states (DOS), total energy, and Fermi level ($E_F$).

The flat band structure from Γ point to Z observed in Fig. 3b is due to the electron confinement and localization in GQDs from the Heisenberg Uncertainty Principle. Because the energy bands that are obtained for the GQDs are of a "flat" type, we use the Γ-point algorithm to choose the k-points at the rest of our work. Algorithms such as the Monkhorst–Pack algorithm [39] for selecting the k-points are not required for GQDs.

As indicated in Fig. 3c, the DOS of GQDs is needle-like and sharp due to the energy levels discretization.

Another phenomenon observed in the electronic properties of some GQDs is the degeneracy of the band of zero-energy states. Mostly in triangular graphene quantum dots (TGQD) with zigzag edges, the electronic density of the highest valence states is localized in the center of the structure. However, in this system, a degenerate shell appears. The energy spectrum near the Fermi level collapses to a degenerate shell at the Fermi level. In TGQD, numerical results, which were predicted first time by the tight-binding model and then confirmed by DFT calculations, show that the zero-energy band states' degeneracy is proportional to the edge size. Edge effects appear only in systems with zigzag edges. Zigzag edges break the symmetry between the two sub-lattices; hence, one expects finite magnetism near those edges. This issue opens up the possibility of designing a strongly correlated electronic system as a function of filling the shell, in analogy to the fractional quantum Hall effect. The number of zero-energy states, $N_{deg}$, is related to the number of edge atoms, $N_{ed}$, through (4), which provide an analytical solution to the edge states[40].

$$N_{deg} = N_{ed} - 1 = N_A - N_B \qquad (4)$$

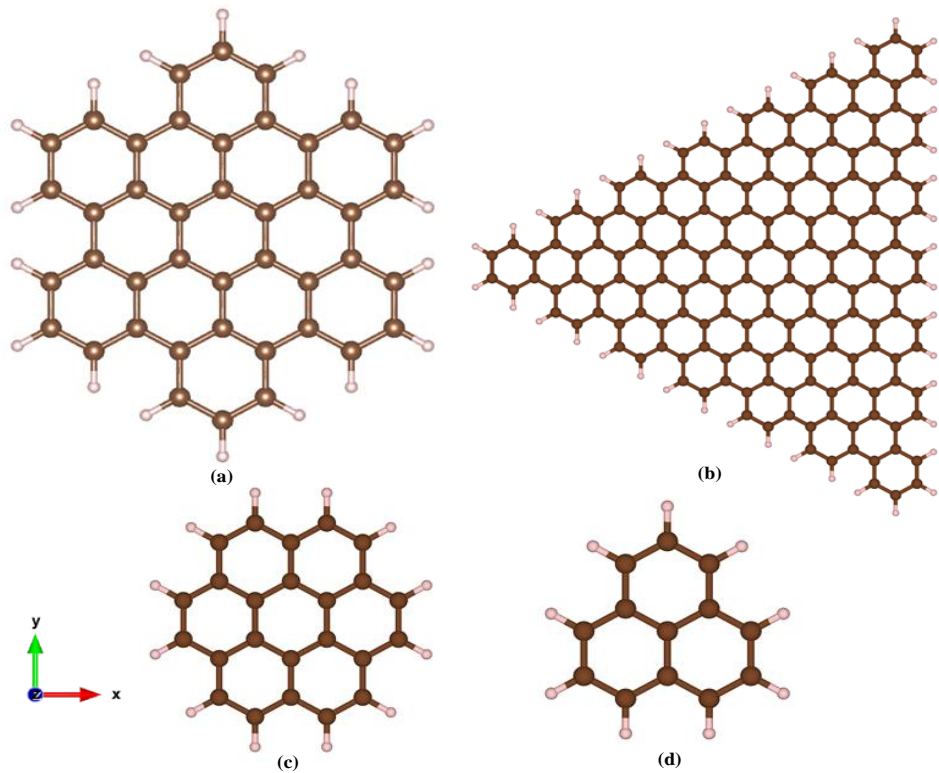

Fig. 1. Schematics of various graphene quantum dots (GQDs) structures. (a) Hexagonal shape with armchair edges, (b) Triangular shape with armchair edge, (c) Hexagonal shape with zigzag edge, and (d) Triangular shape with zigzag edges.

Thus, the degenerate shell is related to the broken sub-lattice symmetry. The origin of this degeneracy is the nonsymmetrical number of sublattices $N_A$ and $N_B$ in the atomic structure of TGQDs that create edge states [21, 40, 41]. Figure 4 illustrates the band structure's difference between the nondegenerate and degenerate GQDs and introduces a method for the determination of the bandgap from the energy band structure in degenerate GQDs [14]. As shown in Fig. 4c, a particular error may be present in the bandgap values determined for the degenerate GQDs. Suppose the smearing parameter is not selected correctly in the DFT calculation code. In that case, the reported results for the bandgap of such GQDs will be incorrect. The small gap between the zero energy levels is considered the bandgap, and the GQD system is incorrectly determined to be metallic.

Table I summarizes the results of the DFT calculations for the electronic properties of the various GQDs. Some of these GQDs, such as $TriC_{13}zzH_9$, $HexC_{19}zzH_{11}$, $TriC_{22}zzH_{12}$, $TriC_{33}zzH_{15}$, and $HexC_{39}zzH_{17}$ are degenerate have zero energy states. Other GQDs listed in Table I are nondegenerate. The problem's size to be solved in DFT calculations is determined by the number of free electrons, $N_e$, in the GQDs structure [42]. This Order-N dependence is the third order, $O(N_e^3)$. Using the information of Table I, columns 1 and 3, we can obtain $N_e$, the number of electrons in each GQD structure. Thus, it is enough to multiply the number of atoms used by the number of free electrons of that atom. The number of free electrons is four for the carbon atom, and the number of free electrons is one for the hydrogen atom. A close examination of the bandgap data presented in Table I shows that all of the quantum dots are direct bandgap semiconductors and have significant bandgaps regardless of their geometric shape, size, and side-edge type. This type bandgap means the formation of confinement-induced energy gaps, which increase linearly with reducing the lateral dimention. Thus, the bandgap energy increases with the decreasing size of the quantum dot.

The total energy of the system includes the interactions between all particles in the system. When considering electrostatic potential, we multiply the charges. Thus, for a minus charge (electrons) and a positive charge (nucleus), the result is always negative, indicating attraction (but for two negative or two positive point charges, the impact is still positive, meaning repulsion). This charges interactions is why all the total energies in Table I are negative. We can also compare the total energy for the graphene sheet and the GQD system. Table I shows that all GQD systems have a total energy value that exceeds that of the graphene sheet. The electric carrier's situation in the graphene sheet prefers a zero-energy reference point since the zero-energy reference point is a (fictitious) system with all the particles (electrons and nuclei) at rest infinitely far away from each other. Conversely, the electrons in GQDs are confined and under the electrostatic potential of the cores.

As shown in Table II, we compared our results with those of other articles and found good agreement between them [20, 21, 43]. In Table II, Ref-a has used density-functional tight-binding (DFTB) calculation via the DFTB+ package with the "3ob parameter set", Ref-b were calculated via configuration interaction (CI) approach with tens of millions of configurations considered, and Ref-c has evaluated via tight-binding model with the 3rd next nearest-neighbor (NNN), and $|t|=2.5$ (eV).

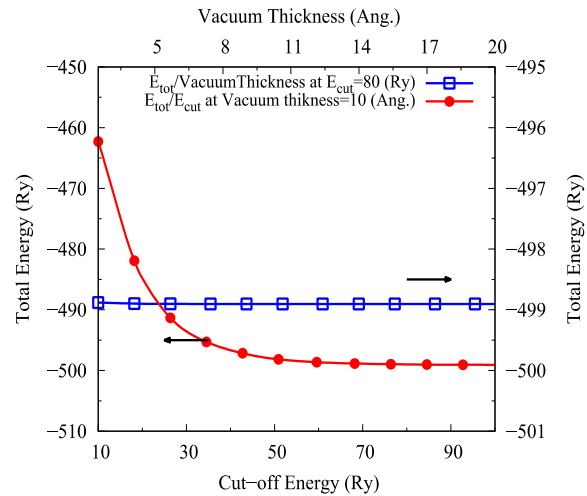

Fig. 2. Minimum energy configuration of the vacuum layer thickness and cut-off energy in density functional theory (DFT) calculations by minimizing the total energy of GQD: Hexagonal $C_{42}acH_{18}$.

### 3.2. Temperature dependence

Table I indicates the bandgap and total energy values at room temperature that were calculated using thDFT methods. Besides, Fig. 5 shows the temperature dependence of the bandgap and total energy for the $C_{42}H_{18}$ GQD. As illustrated in Fig. 5, the electronic properties are almost constant at temperatures up to 900 K. It is observed that for temperatures above 1000 K, both the bandgap and total energy increase. This behavior indicates that the DFT method results for the electronic properties at 0 K are also valid for room temperature conditions. The physical reason for this subject is that the effect of $K_BT$ thermal energy on the free energy functional of GQDs is negligible at room temperature. The total energy is obtained from five different components: (i) external field energy, (ii) ion potential energy, (iii) kinetic energy, (iv) exchange-correlation (xc), and (v) entropy-term. The temperature change has the highest effect on the entropy of the electrons and their kinetic energy. In contrast, the electrostatic potential energy component of ions and the external potential energy applied through the electric field exhibit temperature-independent behavior. However, due to the positive value of kinetic energy and entropy's negative value, the temperature increase is mostly neutralized. Thus, the temperature dependence of the electronic properties is observed only in temperatures above 1000 K.

### 3.3. Tuning the semi-empirical model

Although DFT calculations are highly accurate and do not require the determination of semi-empirical parameters because they are based on first-principles calculations, they need extensive computer resources for GQDs with many atoms. Therefore, we must use the TB model to calculate the electronic properties of large GQDs by tuning the model parameters to match the results of the DFT calculations.

**Table I.** Results of DFT and thermal DFT (thDFT) calculations for graphene and various graphene quantum dots (GQDs).

| System | Size (Å) | Atoms No. | Total Energy (Ry) | | Bandgap (eV) | |
|---|---|---|---|---|---|---|
| | | | Standard DFT | thDFT | Standard DFT | thDFT |
| Graphene | sheet | periodic | $-2.2652\times10^1$ | $-2.2243\times10^1$ | Semimetal | |
| $TriC_{13}zzH_9$ | 6.82 | 22 | $-1.5830\times10^2$ | $-1.5701\times10^2$ | 4.878 | 4.911 |
| $HexC_{19}zzH_{11}$ | 8.32 | 30 | $-2.2890\times10^2$ | $-2.2890\times10^2$ | 3.380 | 3.370 |
| $TriC_{22}zzH_{12}$ | 9.17 | 34 | $-2.6420\times10^2$ | $-2.6500\times10^2$ | 4.410 | 4.485 |
| $HexC_{24}zzH_{12}$ | 9.51 | 36 | $-2.8712\times10^2$ | $-2.8707\times10^2$ | 2.852 | 2.845 |
| $TriC_{33}zzH_{15}$ | 11.60 | 48 | $-3.9290\times10^2$ | $-3.9281\times10^2$ | 4.187 | 4.294 |
| $HexC_{39}zzH_{17}$ | 11.74 | 56 | $-4.6360\times10^2$ | $-4.6351\times10^2$ | 2.311 | 2.324 |
| $HexC_{42}acH_{18}$ | 13.39 | 60 | $-4.9899\times10^2$ | $-4.9800\times10^2$ | 2.300 | 2.328 |
| $TriC_{60}acH_{24}$ | 17.59 | 84 | $-7.1092\times10^2$ | $-7.1078\times10^2$ | 2.174 | 2.174 |
| $HexC_{66}zzH_{20}$ | 13.55 | 86 | $-7.7470\times10^2$ | $-7.7451\times10^2$ | 1.158 | 1.143 |
| $HexC_{96}zzH_{24}$ | 19.04 | 120 | $-1.1211\times10^3$ | $-1.1208\times10^3$ | 1.290 | 1.275 |
| $TriC_{168}acH_{42}$ | 30.19 | 210 | $-1.9620\times10^3$ | $-1.9615\times10^3$ | 1.378 | 1.373 |

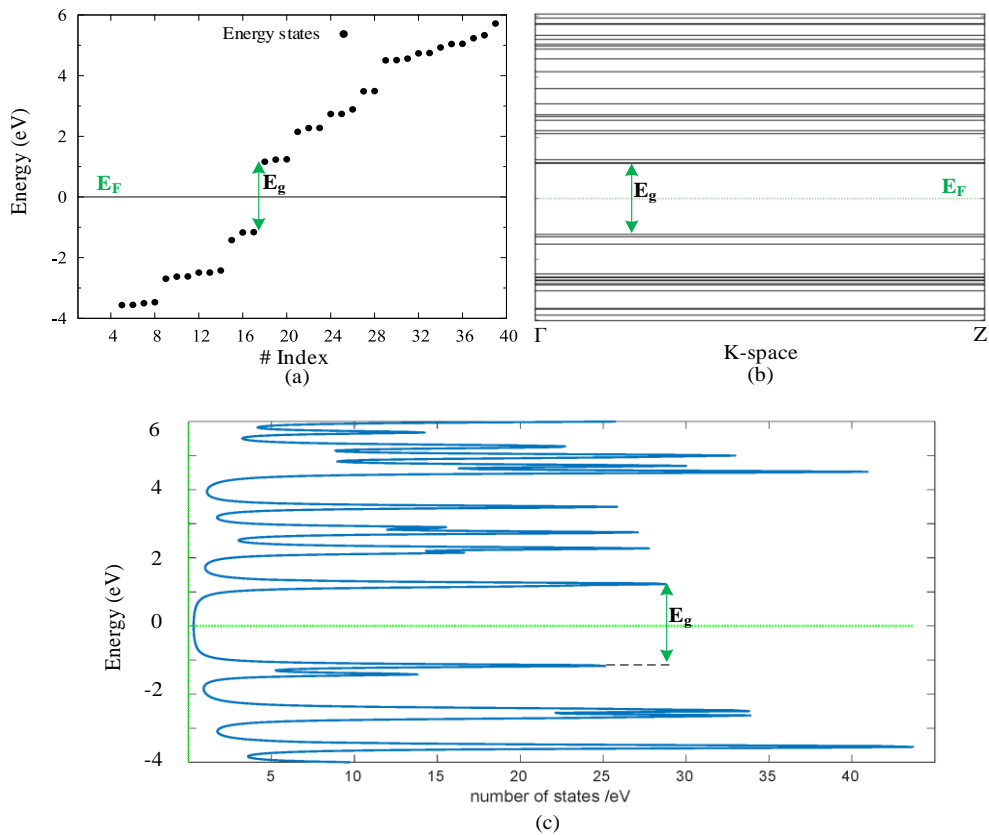

Fig. 3. Electronic properties of the hexagonal $C_{42}acH_{12}$ GQD. (a) molecular energy spectrum (MES), (b) energy band structure from $\Gamma$ point to Z, and (c) density of states (DOS).

**Table II.** Comparing the various GQD bandgaps from our DFT calculations with those from other works using different models, (Ref-a) density-functional tight-binding (DFTB), (Ref-b) configuration interaction (CI), and (Ref-c) tight-binding (TB).

| System | Bandgap (eV) | | | |
|---|---|---|---|---|
| | This work | Ref-a [20] | Ref-b [43] | Ref-c [21] |
| $HexC_{24}zzH_{12}$ | 2.852 | 2.850 | | |
| $HexC_{42}acH_{18}$ | 2.300 | 2.482 | | 2.346 |
| $TriC_{60}acH_{24}$ | 2.174 | 2.202 | 2.230 | |
| $HexC_{96}zzH_{24}$ | 1.290 | 1.310 | | 1.147 |
| $TriC_{168}acH_{42}$ | 1.378 | | | 1.239 |

The results obtained by the TB model depend strongly on the choice of semi-empirical parameters, which is the main bottleneck of TB. However, TB calculations require significantly lower computer resources than DFT calculations. First, we tune the TB model's semi-empirical parameters by matching the DFT results for small GQDs, we generalized this tuned TB model for large GQDs up to thousands of atoms to obtain results with acceptable accuracy using only limited computer resources. The bandgaps of the desired GQDs obtained using both the TB model, DFT, and thDFT calculations are shown in Table III.

**Table III.** The bandgap values for $HexC_{42}acH_{18}$ GQD obtained by the tunned TB model, DFT, and thDFT calculations.

| System | Bandgap (eV) | | |
|---|---|---|---|
| | TB | DFT | thDFT |
| $HexC_{42}acH_{18}$ | 2.2952 | 2.3004 | 2.3283 |

## 4. Conclusions

In this paper, we have elucidated the electronic properties of GQDs using DFT calculations. Calculations using the standard DFT in the ground state, and thDFT were performed. The hexagonal and triangular shapes of GQDs with zigzag and armchair edges were investigated. We compared our results with those other articles with different models and found good agreement between them. It was found that these GQDs are direct bandgap semiconductors and have flat energy band structures from Γ point to Z in the E-k coordinate. The results show that when the A and B sublattices' symmetries are broken, the zero-energy shell is created. Furthermore, we have investigated the temperature dependence of the bandgap and total energy for GQDs in the range from 0 to 1400 K. The calculated results demonstrate that the bandgap and total energy remain constant up to the temperature of 900 K. Therefore, it can be assumed that all DFT calculations in the ground state for extracting the electronic properties obtain results that are valid even for room temperature conditions. Also, we have used the obtained DFT results for tuning the tight-binding model.

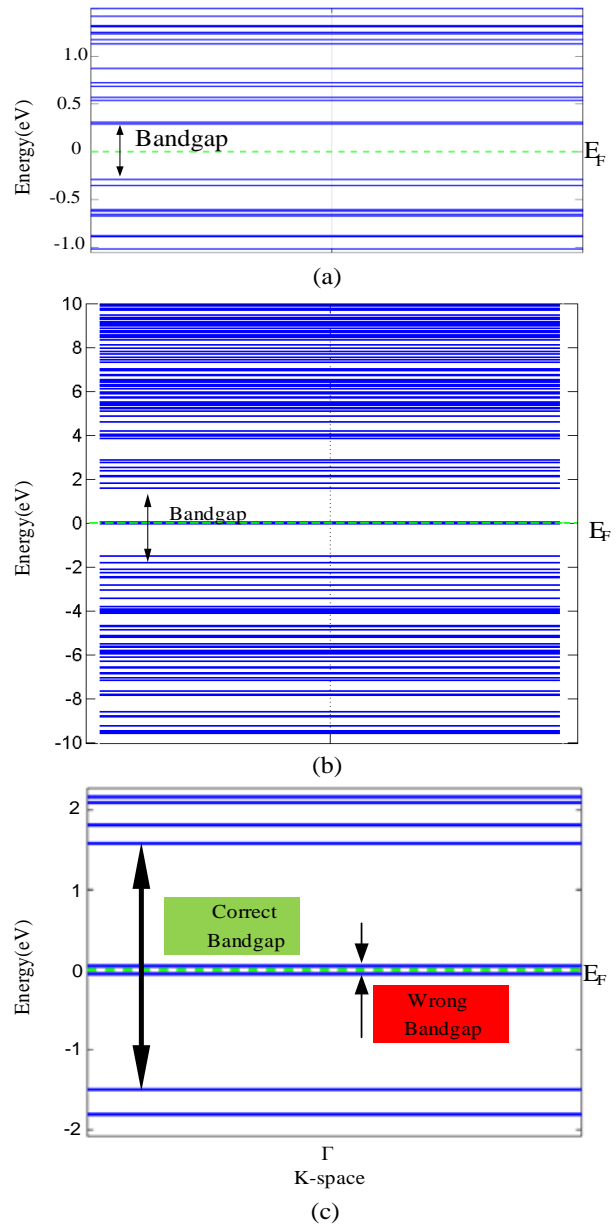

Fig. 4. Energy band degeneracy in the electronic structure calculations of graphene nanodots. (a) system without degeneracy ($C_{66}H_{20}$), (b) system with energy band degeneracy ($C_{39}H_{17}$), and (c) magnified view of the bandgap in the system shown in (b).

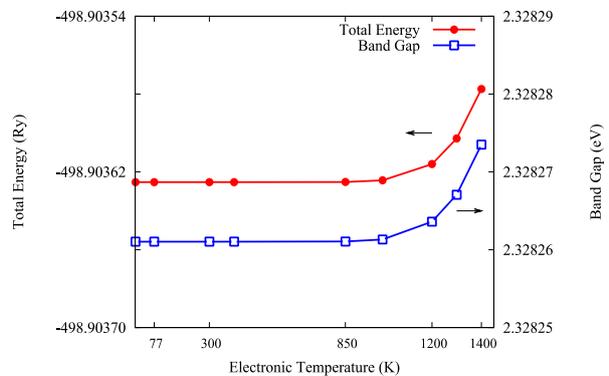

Fig. 5. Temperature dependence of the calculated bandgap and total energy for hexagonal armchair $C_{42}acH_{18}$ GQD obtained by thDFT.